# What drives the charge/orbital ordering in $La_{0.5}Sr_{1.5}MnO_4$: Jahn-Teller distortion versus orbital ordering

Shortened title: JTD versus OO in LSMO

U. Staub[1], V. Scagnoli[1], A. M. Mulders[1], M. Janousch[1], Z. Honda[2], and J. M. Tonnerre[3]

[1] Swiss Light Source, Paul Scherrer Institut, 5232 Villigen PSI, Switzerland

[2] Department of Functional Materials Science, Saitama University, Shimo-Ohkubo 255, Saitama, 338-8570, Japan

[3] Laboratoire de Cristallographie, CNRS, 38042 Grenoble, France

Temperature dependences of a magnetic and an orbital Bragg reflection at the Mn $L_{2,3}$ edges of $La_{0.5}Sr_{1.5}MnO_4$ have been collected. The temperature evolution of the orbital reflection depend strongly on the x-ray energy and reflect the different order parameters for orbitals and Jahn-Teller distortions. The gradual increase of the orbital order parameter indicates that the orbital order and not the Jahn-Teller distortion drives the metal-insulator transition. Correspondingly, the orbital-orbital (super) exchange interaction dominates over the Jahn-Teller strain-field interaction.

PACS: 75.25.+z, 71.30.+h, 75.47.Lx

Mixed valent perovskites show a variety of interesting electronic and magnetic properties such as high-$T_c$ superconductivity, colossal-magneto resistance and/or temperature driven metal-insulator (MI) transitions. In manganites, localization of 3d electrons leads to charge, magnetic and orbital ordering and various models have been proposed to describe the insulating state. At half doping, the Mn ions are found at two distinct sites, which are labeled in a simple charge ordered model by $Mn^{3+}$ and $Mn^{4+}$. At elevated temperatures, these materials show a single site with $Mn^{3.5+}$. Lowering the temperature, a metal-insulator (MI) type transition occurs, and charge order (CO) is combined with an orbital ordering (OO) transition[1] of the $Mn^{3+}$ ions. This simple charge ordering picture has been challenged recently by results obtained from magnetic neutron diffraction,[2] introducing a model where these two Mn sites have equivalent charges and form a dimer, a so called Zener polaron. In addition, recent resonant x-ray scattering experiments at the Mn $K$ edge[3, 4] indicate that the effective charge difference between the two sites is significantly less than one electron charge, though the structural model disagrees with the neutron study.

The $Mn^{3+}$ ion, which resides in an octahedral environment, has a doubly degenerate $e_g$ ground state. This allows orbital ordering associated with a Jahn-Teller distortion (JTD) to occur. However, it is unclear which of the two is the driving force. It has been proposed that it is of electronic origin,[5], 231 (1983)] *e.g.* quadrupole super exchange, or alternatively that it is due to the elastic strain coupled by JTD.[6] Very little is known from an experimental point of view, as most techniques used to investigate orbital ordering phenomena are indirect. Conclusions regarding OO are usually drawn from the detection of the magnetic dipole moments, from the crystallographic distortions (Jahn Teller) or from symmetry considerations.

Single layered manganate $La_{0.5}Sr_{1.5}MnO_4$ exhibits a MI transition at $T_{MI} = T_{CO} = T_{OO} \approx 240$ K, which is followed by an antiferromagnetic transition at $T_N \approx 110$ K. The crystal structure at room temperature is tetragonal with space group I4/mmm and lattice constants $a$=3.86 Å and $c$=12.44 Å. The MI transition causes an increase of the unit cell of type $\sqrt{2}a \times \sqrt{2}a \times c$, which results in two inequivalent Mn sites, as found by neutron diffraction[7]. The magnetic structure shows a magnetic ordering wavevector of (1/4, 1/4, 1/2), leading to a $2\sqrt{2}a \times 2\sqrt{2}a \times 2c$ magnetic unit cell. This is consistent with an orbital ordering with a unit cell of $\sqrt{2}a \times 2\sqrt{2}a \times c$. Further support for orbital ordering was obtained by resonant x-ray scattering at the Mn $K$ edge by the observation of ($h$/4, $h$/4, 0) type reflections.[8]

More recently, resonant soft x-ray scattering at the Mn $L_{2,3}$ edges has been used to investigate the orbital order reflection (1/4, 1/4, 0).[9-11] First principle calculations[12] predict a different photon energy dependence of the scattered intensity for a situation where OO (direct $3d$ state splitting) and JTD ($2p$-$3d$ hybridization) ordering are dominant. Therefore, experiments performed at the $L_{2,3}$ edge allow us to directly probe the associated OO of the $3d$ states as well as the JTD. This is in contrast to the Mn $K$-edge experiments, for which theoretical calculations indicate[13, 14] that the scattered intensity is dominated by the JTD.

The theoretical predictions for the Mn $L_{2,3}$ edges[9, 12, 15] have been compared with recent soft x-ray experiments.[9, 10] However, the assignment of different features in the spectra to different origins remains a challenge. The energy profile at the Mn $L_2$ and $L_3$ edges changes when passing through the magnetic ordering temperature $T_N$. [10, 16] This effect has been interpreted in terms of an enhancement JTD caused by magnetic ordering.[16] In contrast to this, it has recently been shown that a magnetic moment contributes to the orbital reflection below $T_N$.[17] This contribution originates from a minority magnetic phase with ferromagnetic coupling along the $c$-direction. Moreover, the intensity recorded at several x-ray energies show a different temperature dependence for $T_N<T<T_{OO}$, in contrast to the first experimental results.[10]

We present detailed soft x-ray resonant scattering data on the temperature dependence of the orbital and magnetic reflections in $La_{0.5}Sr_{1.5}MnO_4$. The intensity of the OO reflection shows a different temperature dependence for different x-ray energies. This is due to different order parameters for charge ordering and Jahn-Teller distortion. From the gradual change of the orbital order parameter, we conclude that the phase transition is driven by the orbital-orbital interaction and not by the strain fields caused by the JTD.

The single crystal of $La_{0.5}Sr_{1.5}MnO_4$ is the same that used in Ref. [17], where details about the sample preparation are given. Soft x-ray scattering experiments were performed at the SIM beamline of the Swiss Light Source at the Paul Scherrer Institut using the RESOXS endstation.[18] The sample was mounted with the [110] and the [001] directions in the scattering plane, and the orbital (1/4 1/4 0) and the magnetic (1/4 1/4 1/2) reflection were recorded at the Mn $L_{2,3}$ edges. A continuous helium-flow cryostat controlled the sample temperature between 40 and 300 K.

The energy dependence of the orbital (1/4 1/4 0) and magnetic (1/4 1/4 1/2) reflections are compared with one another in Figure 1 for temperatures below $T_N$. The magnetic and orbital reflection are of similar intensity, in contrast to the case of simple perovskite manganate, where the magnetic scattering was found to be much larger than the orbital scattering.[19] Figure 2 shows the temperature dependence of the orbital reflection taken at three distinct energies in the vicinity of both the Mn $L_3$ and $L_2$ edges. These intensities are normalized to that obtained at 170K. The individual curves taken at different energies clearly deviate from one another. These measurements are in contrast to those in Ref. [10] but consistent with those in Ref. [9]. At the Mn $L_{2,3}$ edges, this orbital reflection is believed to originate solely from the asphericity of the Mn $3d$ shell and reflects an atomic electric quadrupole. This is supported by the azimuthal-angle dependence of these intensities taken at the same energies.[17] The azimuthal dependences all have the same form and show the two fold symmetric behavior expected for this charge and orbital ordered phase ($T>T_N$). Therefore, for the single Mn $3d$ quadrupole, an energy independent temperature dependence is expected. In contrast to this expectation, the different temperature evolutions observed indicate different contributions to the scattering. Significant differences are also observed when crossing $T_N$, where the slope of the intensity changes for some energies, whereas for others little or no effect of the magnetic ordering is seen. Below $T_N$, an additional magnetic contribution complicates the analysis of order parameters at the orbital reflection.

To obtain more information on the very unusual temperature dependence of the orbital reflection, we measured the temperature dependence of the magnetic (1/4 1/4 1/2) reflections at different energies (see Figure 3) on the same sample. These curves all lie on top of one another, within the experimental accuracy. This behavior supports the view that there is indeed a single magnetic order parameter for this reflection.

To extract which temperature dependence reflects the order parameter driving the orbital and charge order transition, we have extrapolated the temperature dependences above $T_N$, in the range T=130-200K, down to zero temperature. The square root of the intensities and its extrapolation are shown in Fig. 4. In addition, the JTD order parameter has been extracted from the data of the $K$ edge[20] by normalizing the square root of the intensity at the maximum (~150 K) to the value of one (also shown in Fig. 4). The $K$-edge data are similar in form to the high-energy feature of the $L_3$ and $L_2$ edges. The other features have a significantly different temperature dependence. As calculations have shown that orbital ordering and JTD give a different energy dependence of the (1/4 1/4 0) reflection, we conclude that two different order parameters are present that of the JTD and that of the orbital ordering. Correspondingly, the data labeled with C and F together with the $K$-edge data, reflect the JTD order parameter, and those labeled with A and B are closest to representing the orbital order parameter.

Let us discuss other possible causes for the observed temperature dependence. A structural contribution can be excluded, as it would lead to intensity in the unrotated channel, which is observed neither at the Mn $K$ nor the Mn $L_{2,3}$ edges.[8, 17] Interference with a magnetic contribution is possible below $T_N$, where a magnetic signal was observed with the same wave vector as the orbital signal.[17] Above $T_N$ a very weak signal from the 2-dimensional magnetic scattering may exist.[17] However, such a signal is expected to be much weaker at the $L_2$ edge compared to the $L_3$ edge, which is not observed. Finally, calculations have shown[15] that the energy dependence may strongly depend on the magnetic structure, and we mention the possibility of a different magnetic structure above $T_N$. This is unlikely because there is a loss of coherence, and the local magnetic axis is little affected, as indicated by the observation of 2-dimensional magnetic correlations. Moreover, the detailed calculations cannot explain the observed different temperature evolutions of the different spectral features[15].

A careful analysis of the enhancement of the OO reflection at $T_N$ shows that it cannot solely be described by an additional magnetic contribution. Therefore, the observation that the intensities reflecting the JTD are little affected at $T_N$ is in agreement with the fact that no structural change occurs at $T_N$. On the other hand, the 3d states sensitive to the OO may be affected by the magnetic ordering, as indicated by the behavior of the OO parameter at $T_N$.

The difference in the order parameter of the JTD and the electronic orbital ordering may be caused by the different time scales of the fluctuations. The JTD fluctuations involve the masses of the Mn and O ions, hence its fluctuations are slower than those of the electronic charge hopping process. Recently, inelastic neutron scattering has shown that the orbital fluctuations are much faster than those of the magnetic dipole moments.[21] In addition, different order parameters for charge order and the associated structural distortion have been proposed for $NdNiO_3$[22] and $Yb_4As_3$[23] and, it has been argued that this could be caused by the different timescales of the fluctuations. The order parameter for the OO exhibit a gradual increase for decreasing temperatures, in contrast to the order parameter of the JTD, which shows a turnover at approximately 150K. This turnover at 150 K indicates that the JTD is not the primary order parameter, as the primary order parameter is expected to continue increasing down to zero temperature. This turnover could be caused by the rigid Coulomb repulsion of the contracting lattice which prevents a further distortion of the octahedra. The further increase of the OO order parameter indicates that the coupling between the OO and the JTD is not strong enough and the system can still gain energy enhancing the orbital ordering. This fact that the charge order reflection observed at the *K*-edge [20] has a similar temperature dependence as the JTD indicates that the OO is the primary order parameter. Therefore, it is the orbital – orbital (super) exchange interaction which is driving the MI transition.

In summary, we have determined in detail the temperature dependence of the magnetic (1/4 1/4 1/2) and the orbital (1/4 1/4 0) reflection at different energies in the vicinity of the $L_{2,3}$ edges. The temperature dependences of the magnetic reflection are similar, whereas they depend strongly on the energy for the orbital reflection. We interpret this behavior as being caused by different order parameters for orbital order and the Jahn-Teller distortion. The orbital order parameter exhibits a gradual temperature dependence indicative of the driving force of the metal-insulator transition. Correspondingly, the orbital-orbital (super) exchange interaction dominates the Jahn-Teller strain-field interaction.

We thank the beamline staff of X11MA for its excellent support. This work was supported by the Swiss National Science Foundation and performed at SLS of the Paul Scherrer Institut, Villigen PSI, Switzerland. The work at RIKEN was partially supported by a Grant-in-Aid for Scientific Research from the Japan Society for the Promotion of Science.

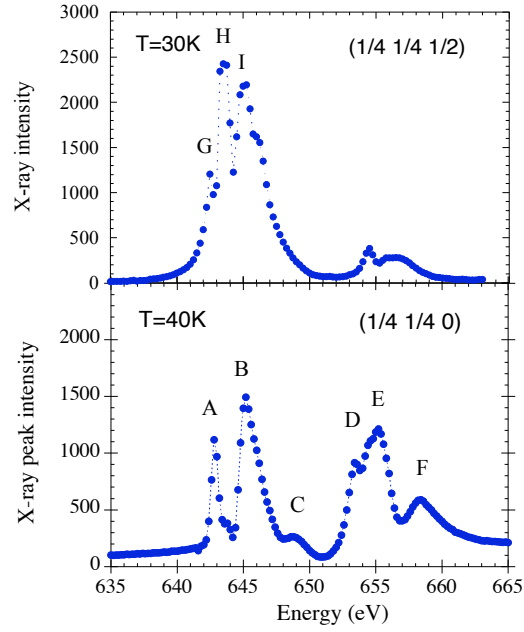

Figure 1 Energy dependence of the magnetic (upper panel) and orbital (lower panel) reflection taken on the same $La_{0.5}Sr_{1.5}MnO_4$ crystal at the Mn $L_{2,3}$ edges.

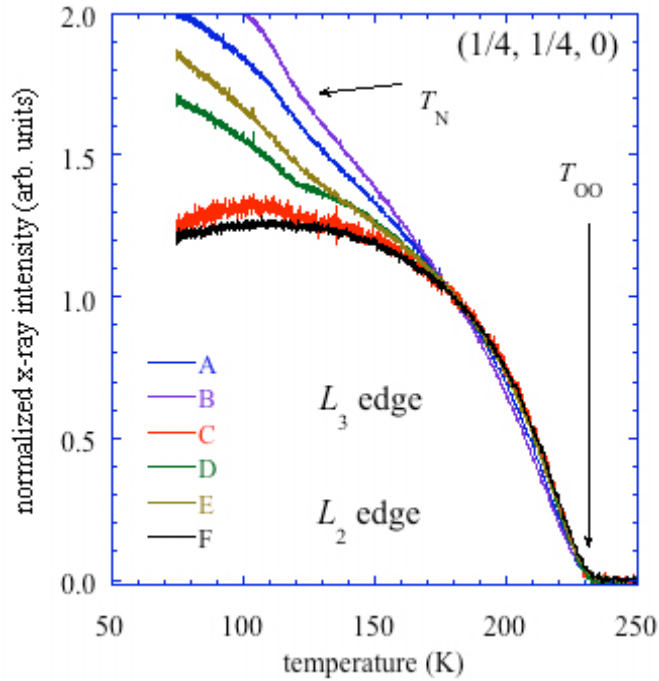

Figure 2 Temperature dependence of the orbital reflection taken at different energies, as labelled in Fig. 1, of the Mn $L_{2,3}$ edges, normalized to the intensity at 170K.

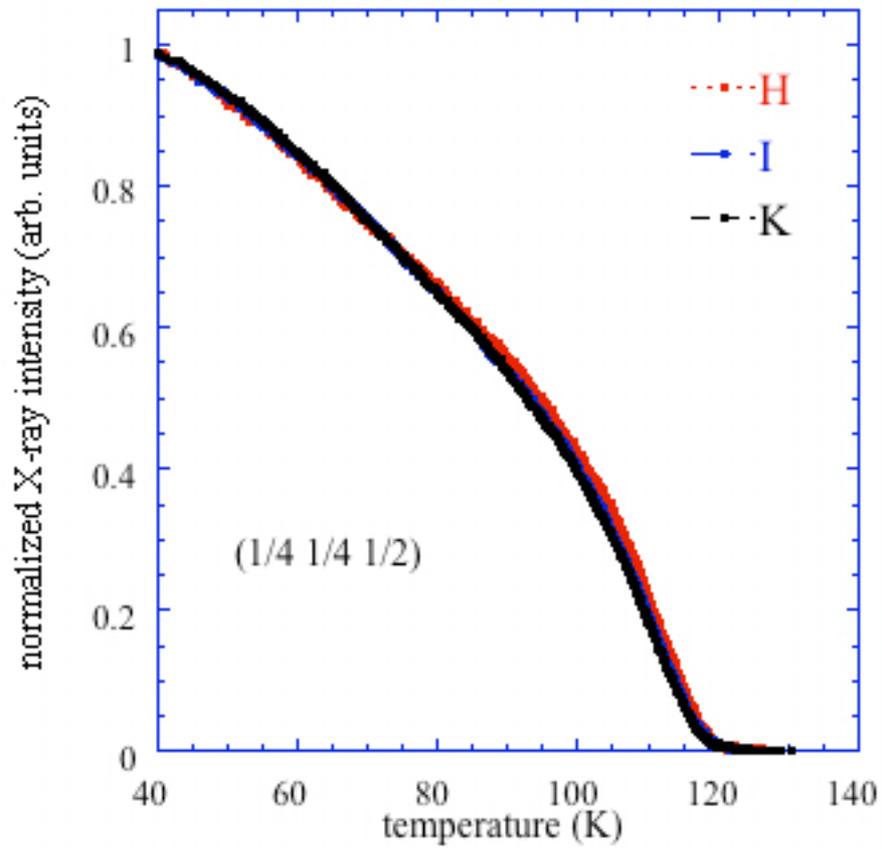

Figure 3 Temperature dependence of the magnetic reflection taken at different energies, as labelled in Fig. 1, of the Mn $L_{2,3}$ edges, normalized to the intensity at 40K.

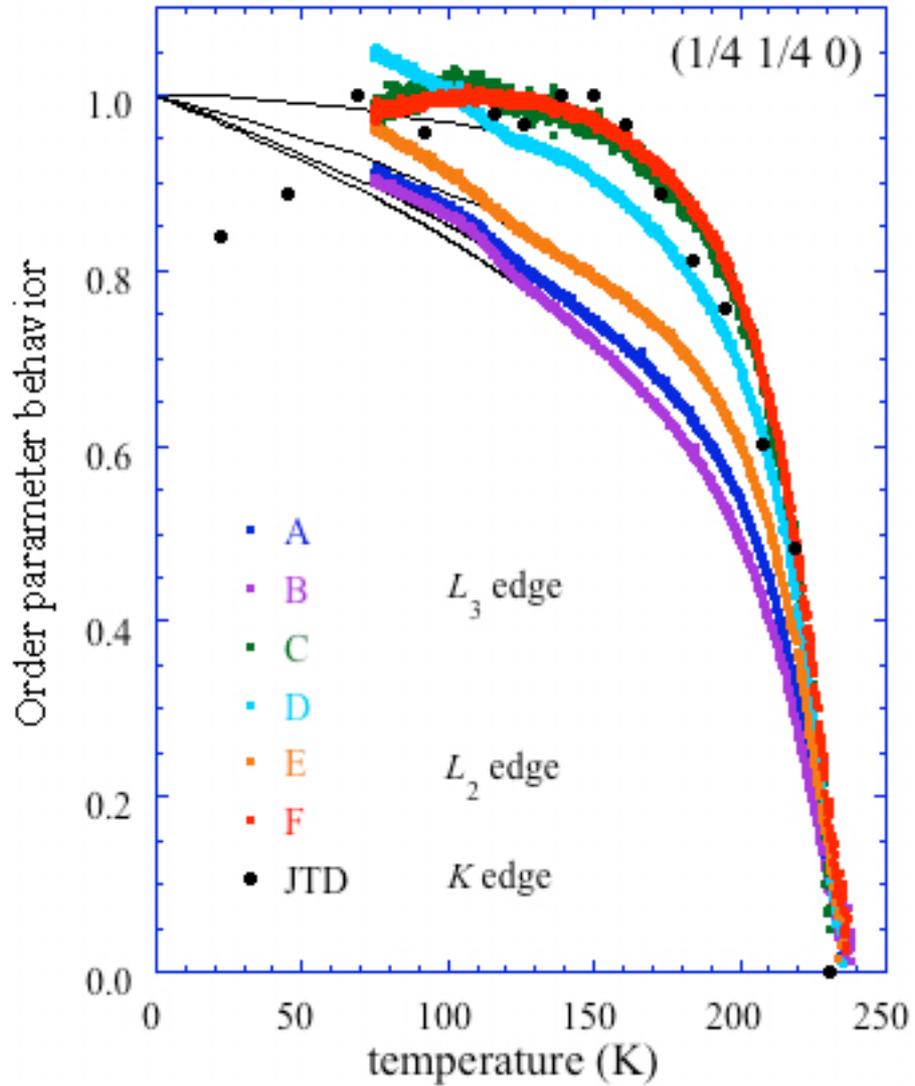

Figure 4 (color online) Temperature dependence of the square root of the intensities taken at different energies of the Mn $L_{2,3}$ edges, normalized to the extrapolated values (lines) at 0 K, reflecting an order parameter behavior. The circles correspond to the intensities of the orbital reflection collected at the Mn $K$-edge.[20]

[1] J. B. Goodenough, Phys. Rev., **100** (1955) 564.
[2] A. Daoud-Aladine, J. Rodríguez-Carvajal, L. Pinsard-Gaudart, M. T. Fernández-Díaz, and A. Revcolevschi, Phys. Rev. Lett., **89** (2002) 097205.

[3]     S. Grenier, J. P. Hill, D. Gibbs, K. J. Thomas, M. v. Zimmermann, C. S. Nelson, V. Kiryukhin, Y. Tokura, Y. Tomioka, D. Casa, T. Gog, and C. Venkataraman, Phys. Rev. B, **69** (2004) 134419.
[4]     J. Herrero-Martín, J. García, G. Subias, J. Blasco, and M. Concepción-Sanchez, Phys. Rev. B, **70** (2004) 024408.
[5]     K. I. Kugel and D. I. Khomskii, Usp. Fiz. Nauk, **136** (1982 [Sov. Phys. Usp. 25, 231 (1983)]) 621.
[6]     D. I. Khomskii and K. I. Kugel, Phys. Rev. B, **67** (2003) 134401.
[7]     B. J. Sternlieb, J. P. Hill, U. C. Wildgruber, G. M. Luke, B. Nachumi, Y. Moritomo, and Y. Tokura, Phys. Rev. Lett., **76** (1996) 2169.
[8]     Y. Murakami, H. Kawada, H. Kawata, M. Tanaka, T. Arima, Y. Moritomo, and Y. Tokura, Phys. Rev. Lett., **80** (1998) 1932.
[9]     S. S. Dhesi, A. Mirone, C. D. Nadai, P. Ohresser, P. Bencok, N. B. Brookes, P. Reutler, A. Revcolevschi, A. Tagliaferri, O. Toulemonde, and G. v. d. Laan, Phys. Rev. Lett., **92** (2004) 56403.
[10]    S. B. Wilkins, P. D. Spencer, P. D. Hatton, S. P. Collins, M. D. Roper, D. Prabhakaran, and A. T. Boothroyd, Phys. Rev. Lett., **91** (2003) 167205.
[11]    S. B. Wilkins, N. Stojic, T. A. W. Beale, N. Binggeli, C. W. M. Castleton, P. Bencok, D. Prabhakaran, A. T. Boothroyd, P. D. Hatton, and M. Altarelli, cond-mat/0410713, (2004).
[12]    C. W. M. Castleton and M. Altarelli, Phys. Rev. B, **62** (2000) 1033.
[13]    P. Benedetti, J. v. d. Brink, E. Pavarini, A. Vigliante, and P. Wochner, Phys. Rev. B, **63** (2001) 060408(R).
[14]    M. Benfatto, Y. Joly, and C. R. Natoli, Phys. Rev. Lett., **83** (1999) 636.
[15]    N. Stojic, N. Binggeli, and M. Altarelli, Phys. Rev. B, **72** (2005) 104108.
[16]    S. B. Wilkins, N. Stojic, T. A. W. Beale, N. Binggeli, C. W. M. Castleton, P. Bencok, D. Prabhakaran, A. T. Boothroyd, P. D. Hatton, and M. Altarelli, Phys. Rev. B, **71** (2005) 245102.
[17]    U. Staub, V. Scagnoli, A. M. Mulders, K. Katsumata, Z. Honda, H. Grimmer, M. Horisberger, and J. M. Tonnerre, Phys. Rev. B, **71** (2005) 214421.
[18]    N. Jaouen, J-M. Tonnerre, G. Kapoujian, P. Taunier, J-P. Roux, D. Raoux, and F. Sirotti, J. Syn. Rad., **11** (2004) 363.
[19]    K. J. Thomas, J. P. Hill, S. Grenier, Y.-J. Kim, P. Abbamonte, L. Venema, A. Rusydi, Y. Tomioka, Y. Tokura, D. F. McMorrow, and M. v. Veenendaal, Phys. Rev. Lett., **92** (2004) 237204.
[20]    Y. Wakabayashi, Y. Murakami, Y. Moritomo, I. Koyama, H. Nakao, T. Kiyama, T. Kimura, Y. Tokura, and N. Wakabayashi, J. Phys. Soc. Jpn., **70** (2001) 1194.
[21]    U. Staub, A. M. Mulders, O. Zaharko, S. Janssen, T. Nakamura, and S. W. Lovesey, Phys. Rev. Lett., **94** (2005) 036408.
[22]    U. Staub, G. I. Meijer, F. Fauth, R. Allenspach, J. G. Bednorz, J. Karpinski, S. M. Kazakov, L. Paolasini, and F. d'Acapito, Phys. Rev. Lett., **88** (2002) 126402.
[23]    U. Staub, M. Shi, C. Schulze-Briese, B. D. Patterson, F. Fauth, E. Dooryhee, L. Soderholm, J. O. Cross, D. Mannix, and A. Ochiai, Phys. Rev. B, **71** (2005) 75115.